\documentclass[aps,prl,twocolumn,superscriptaddress,floatfix]{revtex4}%
\usepackage{graphicx}
\usepackage{dcolumn}
\usepackage{bm}
\usepackage{hyperref}
\usepackage{amssymb}
\usepackage{amsmath}
\usepackage{amsfonts}
\usepackage{amssymb}
\usepackage{color}%
\begin{document}
\title{Resonant control of elastic collisions between $^{23}$Na$^{40}$K molecules and $^{40}$K atoms}
\author{Zhen Su}
\thanks{These authors contributed equally to this work.}
\affiliation{Hefei National Laboratory for Physical Sciences at the Microscale and Department of Modern Physics, University of Science and Technology of China,
Hefei, Anhui 230026, China}
\affiliation{Shanghai Branch, CAS Center for Excellence and Synergetic Innovation Center in Quantum  Information
and Quantum Physics, University of Science and Technology of China, Shanghai 201315, China}
\affiliation{Shanghai Research Center for Quantum Sciences, Shanghai 201315, China}
\author{Huan Yang}
\thanks{These authors contributed equally to this work.}
\affiliation{Hefei National Laboratory for Physical Sciences at the Microscale and Department of Modern Physics, University of Science and Technology of China,
Hefei, Anhui 230026, China}
\affiliation{Shanghai Branch, CAS Center for Excellence and Synergetic Innovation Center in Quantum  Information
and Quantum Physics, University of Science and Technology of China, Shanghai 201315, China}
\affiliation{Shanghai Research Center for Quantum Sciences, Shanghai 201315, China}
\author{Jin Cao}
\affiliation{Hefei National Laboratory for Physical Sciences at the Microscale and Department of Modern Physics, University of Science and Technology of China,
Hefei, Anhui 230026, China}
\affiliation{Shanghai Branch, CAS Center for Excellence and Synergetic Innovation Center in Quantum  Information
and Quantum Physics, University of Science and Technology of China, Shanghai 201315, China}
\affiliation{Shanghai Research Center for Quantum Sciences, Shanghai 201315, China}
\author{Xin-Yao Wang}
\affiliation{Hefei National Laboratory for Physical Sciences at the Microscale and Department of Modern Physics, University of Science and Technology of China,
Hefei, Anhui 230026, China}
\affiliation{Shanghai Branch, CAS Center for Excellence and Synergetic Innovation Center in Quantum  Information
and Quantum Physics, University of Science and Technology of China, Shanghai 201315, China}
\affiliation{Shanghai Research Center for Quantum Sciences, Shanghai 201315, China}
\author{Jun Rui}
\affiliation{Hefei National Laboratory for Physical Sciences at the Microscale and Department of Modern Physics, University of Science and Technology of China,
Hefei, Anhui 230026, China}
\affiliation{Shanghai Branch, CAS Center for Excellence and Synergetic Innovation Center in Quantum  Information
and Quantum Physics, University of Science and Technology of China, Shanghai 201315, China}
\affiliation{Shanghai Research Center for Quantum Sciences, Shanghai 201315, China}
\author{Bo Zhao}
\affiliation{Hefei National Laboratory for Physical Sciences at the Microscale and Department of Modern Physics, University of Science and Technology of China,
Hefei, Anhui 230026, China}
\affiliation{Shanghai Branch, CAS Center for Excellence and Synergetic Innovation Center in Quantum  Information
and Quantum Physics, University of Science and Technology of China, Shanghai 201315, China}
\affiliation{Shanghai Research Center for Quantum Sciences, Shanghai 201315, China}
\author{Jian-Wei Pan}
\affiliation{Hefei National Laboratory for Physical Sciences at the Microscale and Department of Modern Physics, University of Science and Technology of China,
Hefei, Anhui 230026, China}
\affiliation{Shanghai Branch, CAS Center for Excellence and Synergetic Innovation Center in Quantum  Information
and Quantum Physics, University of Science and Technology of China, Shanghai 201315, China}
\affiliation{Shanghai Research Center for Quantum Sciences, Shanghai 201315, China}

\begin{abstract}{We have demonstrated the resonant control of the elastic scattering cross sections in the vicinity of Feshbach resonances between $^{23}$Na$^{40}$K molecules and $^{40}$K atoms by studying the thermalization between them. The elastic scattering cross sections vary by more than two orders of magnitude close to the resonance, and can be well described by an asymmetric Fano profile. The parameters that characterize the magnetically tunable s-wave scattering length are determined from the elastic scattering cross sections. The observation of resonantly controlled elastic scattering cross sections opens up the possibility to study strongly interacting atom-molecule mixtures and improve our understanding of the complex atom-molecule Feshbach resonances.}
\end{abstract}
\maketitle

Feshbach resonances are of great importance to the study of ultracold gases \cite{chin2010}. They occur when the energy of a bound state of a collisional system coincides with that of a scattering state. Magnetic fields can be used to tune the collisional system in and out of resonances if the bound state and the scattering state have different magnetic moments. The magnetically tunable Feshbach resonances have been used as a powerful tool to study strongly interacting quantum gases \cite{Regal2004,Chin2004,Martin2006}, Efimov resonances \cite{Kraemer2006}, and the association of diatomic molecules \cite{Regal2003,Herbig2003}. Resonant control of the s-wave scattering length is  essential to these applications \cite{Tiesinga1993,Inouye1998,Roberts1998,Loftus2002}. In the vicinity of atomic Feshbach resonances, the s-wave scattering length can be described by a simple formula \cite{chin2010}
\begin{equation}
a(B)=a_{\rm{bg}}(1-\frac{\Delta}{B-B_{0}-i \gamma/2}),
\label{eq1}
\end{equation}
where $a_{\rm{bg}}$ denotes the background scattering length, $B_0$ represents the resonance position, $\Delta$ is the resonance width, and $\gamma$ is a decay term that is usually neglected in describing atomic Feshbach resonances. These parameters are usually calculated by coupled-channel calculations using precise Born-Oppenheimer potentials, which are determined by comparing with the experiments.

Feshbach resonances involving ultracold molecules are extremely difficult to understand \cite{Quemener2012}. For the heavy alkali-metal diatomic molecules formed from ultracold atomic gases \cite{Ni2008,Molony2014,Takekoshi2014,Park2015,Guo2016,Rvachov2017,seesselberg2018,LiuL2019,Voges2020}, many rovibrational states of the molecules may contribute to resonant states, and thus it is expected that scattering resonances play crucial roles in ultracold collisions involving molecules. However, the large number of collision channels, the uncertainties of the potential energy surface, and the complexity of the three-body or four-body dynamics make the coupled-channel calculations intractable \cite{Croft2017}. Ultracold collisions involving molecules are usually studied by analyzing the density of short-range chaotic resonant states using the statistical method \cite{Mayle2012,Mayle2013}. The universal loss rate coefficient is one of the main predictions, which is usually compared with the experiments \cite{Ospelkaus2010a,Takekoshi2014,Park2015,Guo2016,Rvachov2017,Voges2020,Ye2018,Bause2021,Gregory2021,Nichols2021}.   However, for atom-molecule collisions, the estimated mean spacing between neighboring states is actually much larger than the temperature of the ultracold gases \cite{Croft2017,Christianen2019pra,Matthew2021} and thus the analysis based on the density of states and the thermal averaging over many resonances is in principle not applicable.

Recently, magnetically tunable atom-molecule Feshbach resonances have been observed in inelastic collisions between singlet ground-state $^{23}$Na$^{40}$K molecules  and $^{40}$K atoms \cite{Yang2019,Wang2021}, and reactive collisions between triplet ground-state $^{23}$Na$^{6}$Li molecules and $^{23}$Na atoms \cite{Son2021}. A detailed analysis of the resonance pattern in collisions of $^{23}$Na$^{40}$K with $^{40}$K suggests that the resonant states associated with these resonances may be the long-range bound states \cite{Wang2021}, but not the short-range chaotic bound states that are usually considered.  The observation of well-isolated individual atom-molecule Feshbach resonances opens up the possibility of controlling the interactions in atom-molecule gases.
However, so far, only the resonantly enhanced losses of the molecules can be studied near the Feshbach resonances. This largely limits the application of the atom-molecule Feshbach resonance, since many important applications of Feshbach resonances require the control of the s-wave scattering length or equivalently, the elastic collision cross sections $\sigma_{\rm{el}}=4\pi|a|^2$.  Due to the large inelastic loss rate coefficients, it is unclear whether tunable elastic collisions can be observed. Although in the collisions between $^{23}$Na$^{6}$Li and $^{23}$Na, a model based on Fabry-Perot interferometer is used to extract the scattering length from the enhanced loss rate coefficients \cite{Son2021}, direct observation of magnetically controlled elastic collisions or the scattering length remains elusive.


In this Letter, we report on the study of resonant control of elastic collisions in the vicinity of a Feshbach resonance between $^{23}$Na$^{40}$K molecules and $^{40}$K atoms.
The $^{23}$Na$^{40}$K molecules and $^{40}$K atoms are both prepared in the maximally spin stretched state.
We study the thermalization between $^{23}$Na$^{40}$K molecules and $^{40}$K atoms, from which the elastic scattering cross sections can be obtained. We find that the elastic scattering cross sections vary by more than two orders of magnitude close to the resonances. The elastic scattering cross sections can be well described by an asymmetric Fano profile, which is a hallmark of Feshbach resonance. The four parameters that characterize a Feshbach resonance are directly obtained from the elastic scattering cross sections. The demonstration of resonantly controlling the s-wave scattering length paves the way towards studying strongly interacting atom-molecule mixtures and improves the understanding of the complex atom-molecule collisions.

Our experiment starts with an ultracold mixture of $^{23}$Na$^{40}$K molecules and $^{40}$K atoms. The experimental procedures for preparing the ultracold mixture have been introduced in previous works \cite{Yang2022,Wang2022}. Briefly, we load laser-cooled $^{23}$Na and $^{40}$K atoms into a cloverleaf magnetic trap to perform evaporative cooling. The atomic mixtures are then loaded into a large-volume three-beam optical dipole trap for further evaporative cooling. At the end of optical evaporative cooling, we create an ultracold mixture containing about $3.0\times10^5$ $^{23}$Na atoms and $2.3\times10^5 $ $^{40}$K atoms. The trap frequencies for $^{40}$K atoms are $(\omega_x,\omega_y,\omega_z)=2 \pi\times (238,71,24)$ Hz. The atomic Feshbach resonance between the $|f,m_{f}\rangle_{\rm{Na}}=|1,1\rangle$ state and the $|f,m_{f}\rangle_{\rm{K}}=|9/2,-9/2\rangle$ state at about 78.3 G is used to form $^{23}$Na$^{40}$K Feshbach molecules. We ramp the magnetic field from 80 G to 77.6 G to create about $2.8\times10^4$ $^{23}$Na$^{40}$K Feshbach molecules. The $^{23}$Na$^{40}$K Feshbach molecules are then transferred to the rovibrational ground state by stimulated Raman adiabatic passage (STIRAP). We prepare the $^{23}$Na$^{40}$K molecules in the maximally polarized hyperfine level $|v,N,m_{i_{\rm{Na}}},m_{i_{\rm{K}}}\rangle=|0,0,-3/2,-4\rangle$ of the rovibrational ground state, where $v$ and $N$ represent the vibrational and rotational quantum numbers, and $m_{i_{\rm{Na}}}$ and $m_{i_{\rm{K}}}$  denote the nuclear spin projections of $^{23}$Na and $^{40}$K respectively. We use the hyperfine level of the excited state $F_1\approx3/2, m_{F_1}\approx-3/2, m_{\rm{K}}\approx-3$ as the intermediate state for the STIRAP \cite{Park2015b,LiuY2020}. The $\sigma^{-}$  polarized pump laser and the $\sigma^{+}$ polarized Stokes laser propagate along the direction of the magnetic field. The efficiency of a round-trip STIRAP is about 50\%. After removing the remaining $^{23}$Na atoms by resonant light pulses, we prepare an ultracold mixture of $^{23}$Na$^{40}$K molecules and $^{40}$K atoms.

We study the elastic collisions between $^{23}$Na$^{40}$K molecules in the $|0,0,-3/2,-4\rangle$ state and $^{40}$K atoms in the $|9/2,-9/2\rangle$ state in the vicinity of an atom-molecule Feshbach resonance located at about 48 G. After the ultracold atom-molecule mixture is prepared, we ramp the magnetic field to a target value in 3 ms. We then apply a 50-$\mu$s weak resonant light pulse to heat the $^{40}$K atoms. After the heating pulse, the temperatures of the $^{40}$K atoms and the $^{23}$Na$^{40}$K molecules are about 500 nK and 300 nK respectively. The heating light pulse also reduces the atom numbers by a factor of about 2. The $^{23}$Na$^{40}$K molecules and the $^{40}$K atoms are held at the target magnetic field for a short time. The uncertainty of the magnetic field is about 200 mG. We then remove the $^{40}$K atoms by a strong resonant light pulse and ramp the magnetic field back to 77.6 G. The $^{23}$Na$^{40}$K molecules are transferred back to the Feshbach state and are detected by absorption imaging along the z direction.

\begin{figure}[ptb]
\centering
\includegraphics[width=8cm]{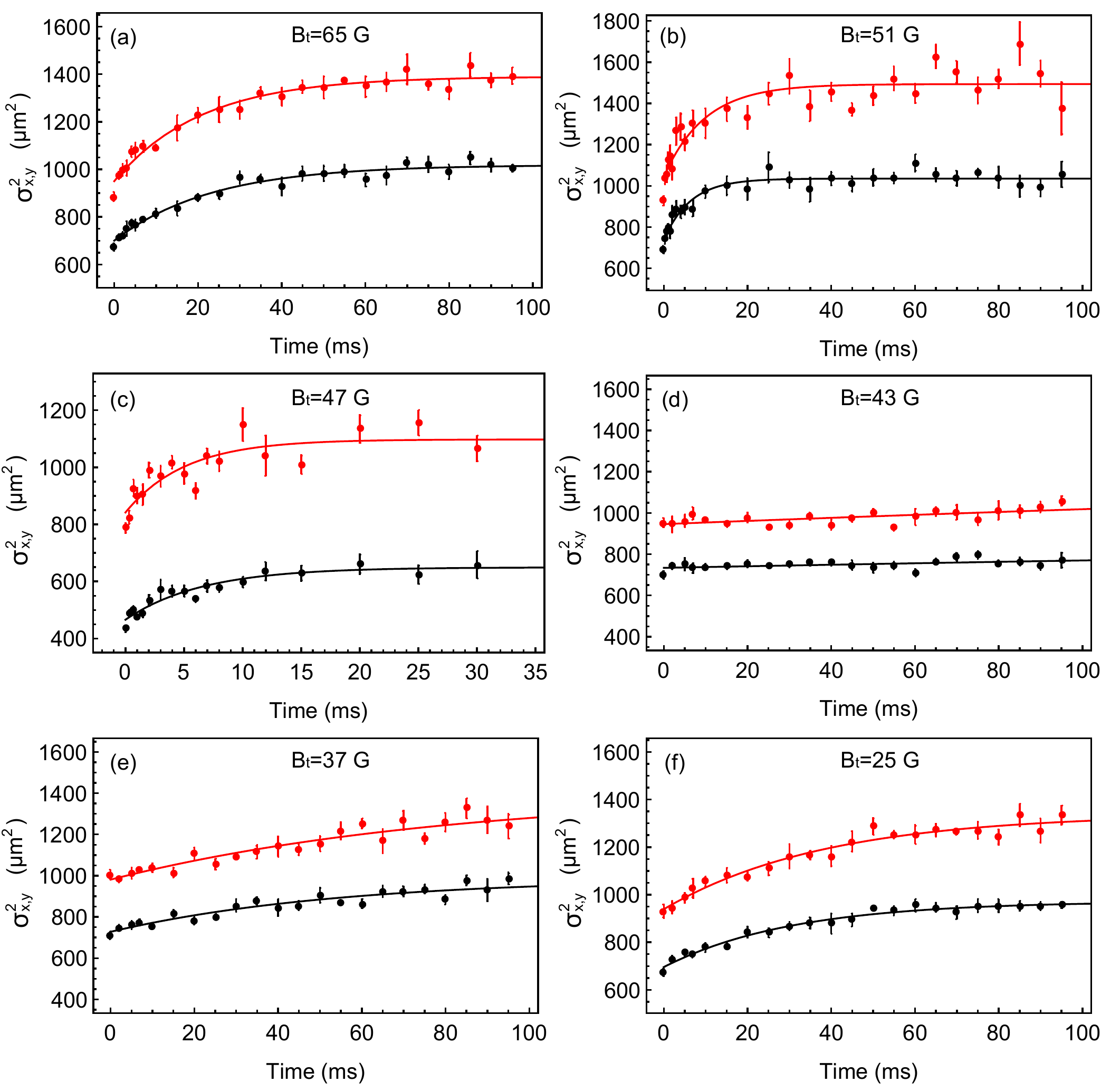}\caption{ The thermalization of the $^{23}$Na$^{40}$K molecules following the heating of the $^{40}$K atoms. The squares of the molecular cloud sizes $\sigma_{x}^2$ (red) and $\sigma_{y}^2$ (black) are plotted as a function of the hold time at various target magnetic fields. The molecular cloud sizes are extracted from the absorption images taken after a time of flight of 3-4 ms. The increases of cloud sizes are due to the elastic collisions between the $^{23}$Na$^{40}$K molecules and the $^{40}$K atoms. The solid lines are the fits of the data points to exponential functions  $\sigma_{x,y}^2=A_{x,y} e^{-\Gamma_{x,y} t}+B_{x,y}$ where the parameters $A, B$ and $\Gamma$ are the fitting parameters. For the magnetic fields close to 43 G at which the thermalization is very weak, the parameters $B$ are fixed. The mean thermalization rate $\Gamma_{\rm{th}}=(\Gamma_{x}+\Gamma_{y})/2$ is used to calculate the elastic scattering cross section.  Error bars represent the standard error of the mean.}
\label{fig1}%
\end{figure}

During the hold time at the target magnetic field, the temperature of the $^{23}$Na$^{40}$K molecules will increase due to elastic collisions with the $^{40}$K atoms. Since the number of $^{40}$K atoms is about one order of magnitude larger than that of the molecules, we may assume that the temperature of the $^{40}$K atoms is constant and the temperature of the $^{23}$Na$^{40}$K molecules changes according to $T(t)=T_f-(T_f-T_i)e^{-\Gamma_{\rm{th}} t}$, where $T_i$ and $T_f$ represent the initial and final temperatures respectively, and $\Gamma_{\rm{th}}$ is the thermalization rate.
The thermalization rate can be extracted from the time evolution of the molecule cloud sizes. For a Maxwell-Boltzmann distribution, the column density distributions obtained from the absorption image can be fit to two-dimensional Gaussian functions $\propto e^{-x^2/(2\sigma_x^2)-y^2/(2\sigma_y^2)}$, where the cloud sizes along the $x$ and $y$ directions can be described by $\sigma_{x,y}^2(t)=k_{\rm{B}}T(t)/m_{\rm{NaK}}(\tau_{\rm{tof}}^2+1/\omega_{x,y}^2)$, with $\tau_{\rm{tof}}$ being the time of flight.

\begin{figure}[ptb]
\centering
\includegraphics[width=8cm]{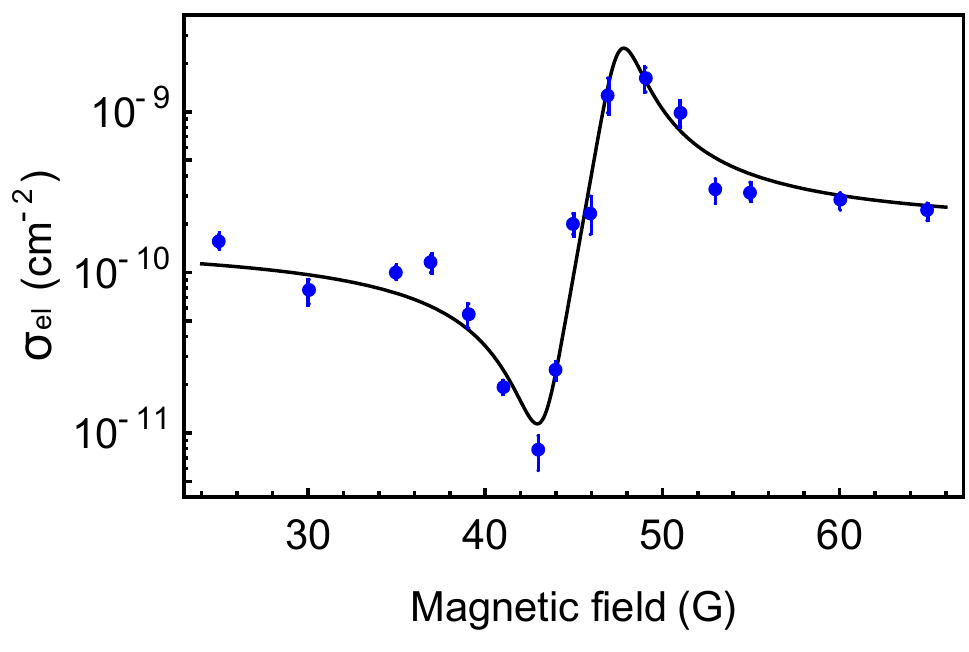}\caption{The elastic scattering cross sections, calculated from the thermalization rate, are plotted as a function of the magnetic fields. The solid lines are the fit using the formula $\sigma_{\rm{el}}=4\pi(\alpha^2+\beta^2)$, where the scattering $a=\alpha-i\beta$ is given by Eq. \ref{eq1}. The parameters obtained from the fit are $a_{\rm{bg}}=-692(36) a_0$, $\Delta=-4.3(4)$ G, $B_0=47.5(3)$ G and $\gamma=2.4(4) $ G. Error bars represent the standard error. }
\label{fig2}%
\end{figure}

The squares of the molecule cloud sizes $\sigma_{x,y}^2$ extracted from the absorption images are shown in Fig. \ref{fig1} as a function of the hold time for several target magnetic fields. The increases of the cloud sizes along both the $x$ and $y$ directions versus the hold time reflect the thermalization due to elastic collisions. It can be clearly seen that the thermalization rate changes with the magnetic fields. The thermalization is very fast at 47 G, which indicates that the elastic collisions are resonantly enhanced. However, at 43 G the thermalization is largely suppressed. This indicates that the elastic scattering cross section has a minimum at about 43 G. Note that at 48 G, it is difficult to determine the thermalization rate due to the strong inelastic losses. We fit the molecule cloud sizes using an exponential function $\sigma_{x,y}^2=A_{x,y} e^{-\Gamma_{x,y} t}+B_{x,y}$ where the parameters $A, B$ and $\Gamma$ are the fitting parameters. For the magnetic fields close to 43 G, the thermalization is very weak, and thus we set the parameter $B$ to the mean value of the fit results obtained at other magnetic fields to obtain a stable fit.

We determine the elastic scattering cross section from the measured thermalization rate using the formula \cite{Mosk2001,Son2020,Tobias2020} $\Gamma_{\rm{th}}= n_{\rm{ov}}\sigma_{\rm{el}} v_{\rm{rel}}/(3/\xi)$, where $n_{\rm{ov}}=(N_{\rm{NaK}}+N_{\rm{K}})[(\frac{2\pi k_{\rm{B}} T_{\rm{K}}}{m_{\rm{K}} \bar\omega_{\rm{K}}^2})(1+\frac{m_{\rm{K}}T_{\rm{NaK}}}{m_{\rm{NaK}} T_{\rm{K}} \gamma_t^2} )]^{-\frac{3}{2}}$
is the overlap density with $\gamma_t=0.79$ being the ratio between the trap frequencies
for $^{23}$Na$^{40}$K molecules and $^{40}$K atoms, $v_{\rm{rel}}=\sqrt{(8 k_{\rm{B}}/\pi)(T_{\rm{K}}/m_{\rm{K}}+T_{\rm{NaK}}/m_{\rm{NaK}})}$
is the relative mean velocity, and the parameter $\xi=4m_{\rm{NaK}}m_{\rm{K}}/(m_{\rm{NaK}}+m_{\rm{K}})^2$ represents the mass effect. The elastic scattering cross sections determined in this way are shown in Fig. \ref{fig2} as a function of the magnetic fields, where the mean value $\Gamma_{\rm{th}}=(\Gamma_{x}+\Gamma_{y})/2$ is used in the calculation.

The elastic scattering cross sections can be changed by more than two orders of magnitude via varying the magnetic field. The elastic scattering cross section shows a clear asymmetric Fano line shape \cite{Fano1961, chin2010}, which is a hallmark of Feshbach resonance. The minimum of the elastic scattering cross section is caused by the Fano interference between the background scattering amplitude in the open channel and the resonant scattering amplitude originating from the bound state in the closed channel. It represents the quantum control of the elastic atom-molecule collisions via quantum interference.

\begin{figure}[ptb]
\centering
\includegraphics[width=8cm]{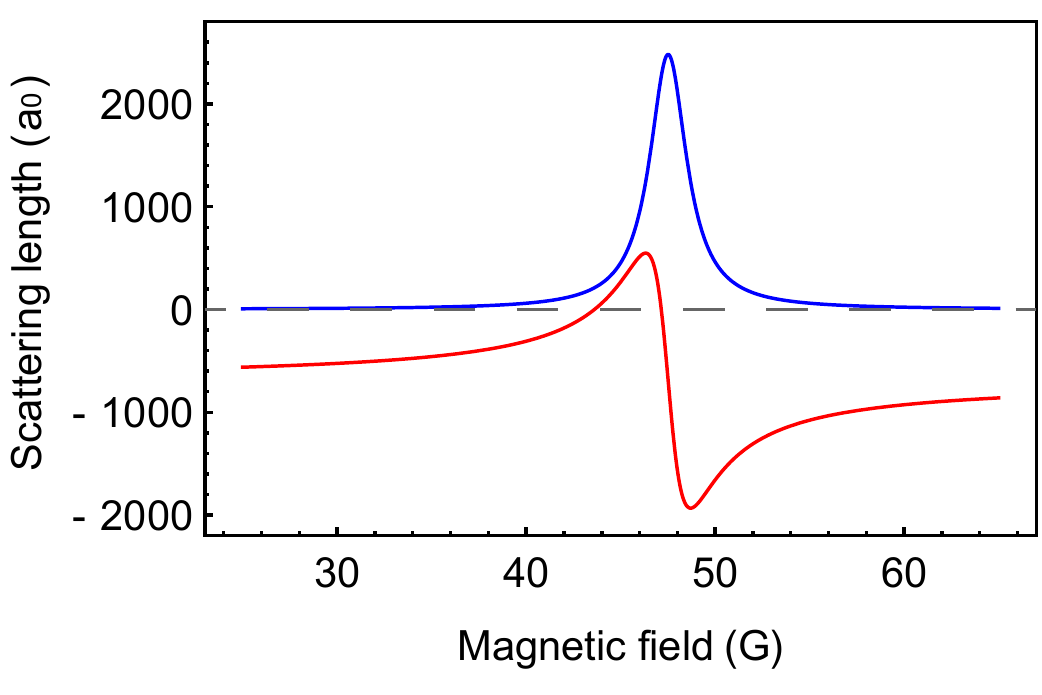}\caption{The real (red) and imaginary (blue) parts of the scattering length are plotted as a function of the magnetic field. The scattering length is plotted according to Eq. \ref{eq1}, with the parameters determined from the elastic scattering cross sections. }
\label{fig3}%
\end{figure}

Although there is no theoretical model that can calculate the elastic scattering cross sections, we may still use Eq. \ref{eq1} to describe the s-wave scattering length  in the vicinity of the atom-molecule Feshbach resonance. In this case, the s-wave elastic scattering cross section may be described by $\sigma_{\rm{el}}=4\pi( \alpha^2+\beta^2)$, where the complex scattering length $a=\alpha-i\beta$ with $\alpha$ and $\beta$ real numbers. When using this simple analytic model, we have neglected thermal average effects. The thermal average will be important when the scattering length is comparable with the de Broglie wavelength, which is about 3000 $a_0$. Therefore we expect that the thermal average mainly affects the accuracy of the resonance position. We fit the measured elastic scattering cross sections using this analytic formula, where $a_{\rm{bg}}, \Delta, B_0$ and $\gamma$ are the fitting parameters.

The fit can only give the absolute value of the background scattering length, but not its sign. The sign of the background scattering length can be determined as follows. Since the minimum of the scattering cross section is below the resonance position, there are two possibilities. If the bound state is on the low field side of the resonance, the background scattering length is negative. On the contrary, the background scattering length is positive. Following the arguments in Ref. \cite{Wang2021}, since both the molecules and atoms are in the maximally polarized state, the triatomic bound states accounting for this resonance must have a nonzero total spin-free rotational angular momentum $J\geq1$, and the bound state must approach the scattering state from the low field side of the resonance. Therefore, the background scattering length is negative. We obtain $a_{\rm{bg}}=-692(36) a_0$, $\Delta=-4.3 (4)$ G, $B_0=47.5 (3)$ G and $\gamma=2.4 (4) $ G. The magnetically tunable scattering length is thus determined. The real and imaginary parts of the scattering length are plotted in Fig.  \ref{fig3}.

\begin{figure}[ptb]
\centering
\includegraphics[width=8cm]{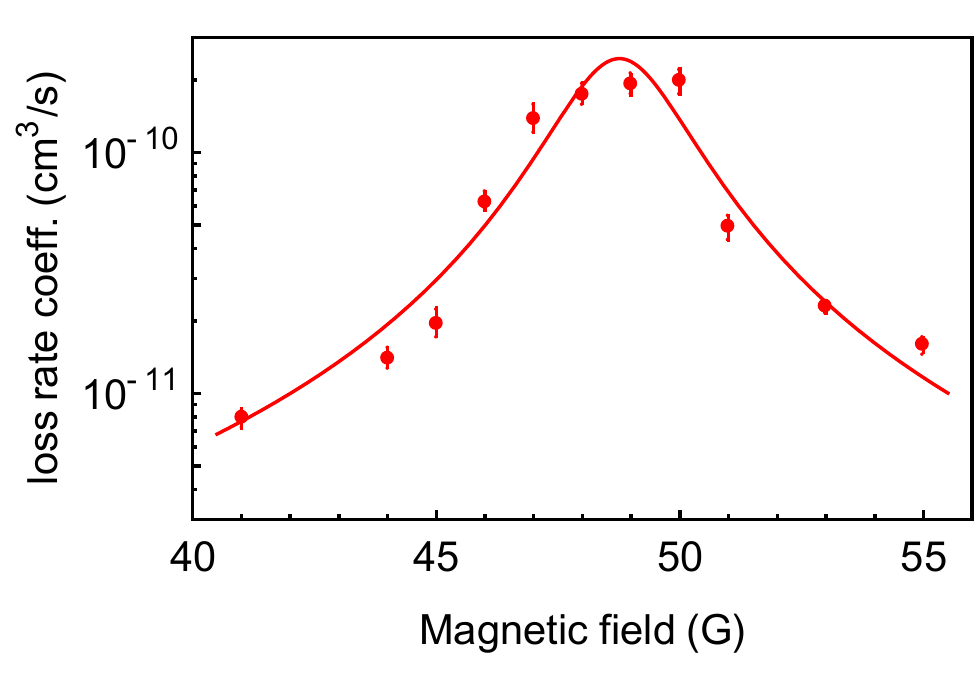}\caption{The loss rate coefficients obtained by measuring the losses of the $^{23}$Na$^{40}$K molecules in the atom-molecule mixture. The solid line is a Lorentzian fit to the data. Error bars represent the standard error. }
\label{fig4}%
\end{figure}

The strong coupling between the bound state and the scattering state causes avoided crossing, and thus the resonance position is different from the intersection between the bare bound state and the scattering state. The intersection between the bare states may be determined using the formula \cite{Julienne2006}
$B_c=B_0-\Delta r_{\rm{bg}}(1-r_{\rm{bg}})/[1+(1-r_{\rm{bg}})^2], $
where $r_{\rm{bg}}=a_{\rm{bg}}/\bar a$ with $\bar a=74a_0$ the mean scattering length. We obtain $B_c=43.7 $ G, which is about 4 G lower than the resonance position.  This value is very close to the narrow resonances at about 43.4 G in the collision channels $|0,0,-1/2,-4\rangle$+ $|9/2,-9/2\rangle$ and $|0,0,1/2,-4\rangle$+$|9/2,-9/2\rangle$ \cite{Wang2021}. This indicates that the bare bound states for these resonances may bunch together, even if the projections of the total angular momentum is different. Therefore, the determination of the bare bound state for broad resonances is important to understand the resonance pattern.  The resonance pattern may be used to estimate the relative moment between the bound state and the scattering state.

It is interesting to compare the elastic collisions with the inelastic collisions. To this end, we measure the loss rate coefficients of the $^{23}$Na$^{40}$K molecules in the atom-molecule mixtures. After ramping the magnetic field to a target value, we monitor the time evolution of the $^{23}$Na$^{40}$K molecules. Assuming the loss of the $^{23}$Na$^{40}$K molecules is caused by two-body inelastic collisions with the $^{40}$K atoms, the loss rate coefficients are given by the ratio between the measured decay rate and the density of $^{40}$K atoms. The measured loss rate coefficients are shown in Fig. \ref{fig4}. The loss rate coefficients are fit to a Lorentz function. The fit also gives the resonance position $B^{\rm{in}}_{0}=48.8(3)$ G and the FWHW width $\gamma^{\rm{in}}=2.8(5)$G. These values are very close to the values determined from the elastic collision cross sections.

In summary, we have demonstrated the resonant control of elastic collisions between $^{23}$Na$^{40}$K molecules  and $^{40}$K atoms. The observation of the magnetic tunable elastic scattering cross sections will help to improve our understanding of the complex atom-molecule Feshbach resonances. In previous theoretical works, different theoretical models are developed to calculate the loss rate coefficients and compare them with experiments. The tunable elastic scattering cross sections near the Feshbach resonances thus provide another important quantity that can be compared with the theory. The elastic scattering cross sections may also be useful in determining the short-range parameters required in the multichannel quantum-defect model for anisotropic interactions \cite{Gao2020}.

The direct observation of the elastic scattering cross sections or the scattering length opens up the possibility of studying strongly interacting atom-molecule mixtures. Recently, ultracold mixtures of molecules and atoms have been proposed to study a new type of impurity problem, i.e., angulon \cite{Schmidt2015,Schmidt2016}, which is a counterpart of the polaron. The atom-molecule Feshbach resonances may be useful since it can provide the strong interaction that is necessary in studying the impurity problem in ultracold gases \cite{Schirotzek2009,Kohstall2012,Hu2016,Arlt2016}. Besides studying strongly interacting atom-molecule mixtures, the universal binding energies of the triatomic bound states close to the resonance position may be determined from the scattering length, which can be compared with the binding energies obtained from the radio-frequency association spectroscopy \cite{Yang2022}.

\begin{acknowledgments}
This work was supported by the National  Key R\&D Program of China (under Grant No.
2018YFA0306502), the National Natural Science Foundation of China (under Grant Nos. 11521063, 11904355),  the
Chinese Academy of Sciences, the Anhui Initiative in Quantum Information Technologies, the Shanghai Municipal Science and Technology Major Project (Grant No.2019SHZDZX01), the Shanghai Rising-Star Program (Grant No. 20QA1410000).
\end{acknowledgments}


\end{document}